\documentclass[aps,prd,twocolumn,nofootinbib]{revtex4}

\usepackage{graphicx,color}

\usepackage[colorlinks=true, linkcolor=blue, citecolor=blue, urlcolor=blue]{hyperref}

\begin{document}

\title{Analysis of the Pion Electromagnetic Form Factor with Next-to-Next-to-Leading Order QCD Corrections}

\author{Sheng-Quan Wang$^{1}$}
\email[email:]{sqwang@alu.cqu.edu.cn}

\author{Zuo-Fen Liao$^{1}$}

\author{Jian-Ming Shen$^{2}$}
\email[email:]{shenjm@hnu.edu.cn}

\author{Hua Zhou$^{3}$}
\email[email:]{zhouhua@swust.edu.cn}

\author{Jia-Wei Zhang$^{4}$}
\email[email:]{jwzhang@cqust.edu.cn}

\author{Jiang Yan$^5$}
\email[email:]{yjiang@itp.ac.cn}

\author{Xing-Gang Wu$^6$}
\email[email:]{wuxg@cqu.edu.cn}

\author{Leonardo Di Giustino$^{7,8}$}
\email[email:]{leonardo.digiustino@uninsubria.it}

\address{$^1$Department of Physics, Guizhou Minzu University, Guiyang 550025, P.R. China}
\address{$^2$School of Physics and Electronics, Hunan University, Changsha 410082, P.R. China}
\address{$^3$School of Mathematics and Physics, Southwest University of Science and Technology, Mianyang 621010, P.R. China}
\address{$^4$Department of Physics, Chongqing University of Science and Technology, Chongqing 401331, P.R. China}
\address{$^5$Institute of Theoretical Physics, Chinese Academy of Sciences, Beijing 100190, P. R. China}
\address{$^6$Department of Physics, Chongqing University, Chongqing 401331, P.R. China}
\address{$^7$Department of Science and High Technology, University of Insubria, via Valleggio 11, I-22100, Como, Italy}
\address{$^8$INFN, Sezione di Milano-Bicocca, 20126 Milano, Italy}

\date{\today}

\begin{abstract}
NNLO QCD corrections for the pion electromagnetic form factor at large momentum transfer have been recently performed in [Phys. Rev. Lett. 132, 201901 (2024); Phys. Rev. Lett. 134, 221901 (2025)], revealing that the NLO and NNLO contributions are positive and sizeable. Unfortunately, these predictions have been obtained using the conventional scale-setting method and thus they are plagued by large renormalization scale ambiguities. In this paper, we analyze the pion electromagnetic form factor at NNLO by applying the Principle of Maximum Conformality (PMC), which is introduced with the aim of resolving renormalization scheme and scale ambiguities. By applying the PMC, a more precise perturbative QCD (pQCD) prediction for the pion EMFF \(Q^2F_\pi(Q^2)\) without conventional renormalization scale ambiguity can be achieved. This improved pQCD prediction is highly beneficial for the precise determination of the pion light-cone distribution amplitude. We then conduct a comprehensive comparison between theoretical predictions and experimental measurements of the pion EMFF \(Q^2F_\pi(Q^2)\).
\end{abstract}

\maketitle

\section{Introduction}

The $\pi$ meson is of great significance in the progress of Quantum Chromodynamics (QCD). The pion electromagnetic form factor (EMFF), denoted as \(F_\pi(Q^2)\), acts as a vital tool for exploring QCD dynamics. It serves to connect the perturbative and non-perturbative domains of the strong interaction. It is defined by the matrix element of the electromagnetic current,
\begin{eqnarray}
\langle \pi^+(p')|J_{\rm em}^\mu|\pi^+(p)\rangle = F_\pi(Q^2) (p^\mu +p'^\mu),
\label{EQ:FQ}
\end{eqnarray}
where $p$ and $p'$ are four-momenta carried by the initial and final pion states, respectively, and the squared momentum transfer $Q^2=-(p-p')^2$. The electromagnetic current is given by $J_{\rm em}^\mu=\sum_q\,e_q\,\bar{q}\,\gamma^{\mu}\,q$ with $e_u=+2/3$ and $e_d=-1/3$ being the electric charges of the up and down quarks.

Experimentally, the pion EMFF has been extensively investigated. Early experimental studies on the pion EMFF up to $Q^2=10$ GeV$^2$ can be traced back to the Cornell experiment~\cite{Bebek:1974ww,Bebek:1976qm,Bebek:1977pe}. The DESY has reported experimental measurements of the pion EMFF since 1977~\cite{Ackermann:1977rp,Brauel:1979zk}. Subsequent experimental results for the pion EMFF at low and intermediate momentum transfers have been obtained at Fermilab~\cite{Dally:1981ur,Dally:1982zk}, CERN~\cite{Amendolia:1984nz,NA7:1986vav} and JLab~\cite{JeffersonLabFpi:2000nlc,JeffersonLabFpi:2007vir,JeffersonLabFpi-2:2006ysh,JeffersonLab:2008gyl,JeffersonLab:2008jve}. The upgraded JLab plans to make future measurements for the pion EMFF up to $Q^2=6$ GeV$^2$~\cite{Dudek:2012vr,Arrington:2021alx}. The determination of the pion EMFF at higher momentum transfers with high precision is one of the key targets for future experiments at the electron-ion collider (EIC)~\cite{AbdulKhalek:2021gbh,Arrington:2021biu} and at the electron-ion collider in China (EicC)~~\cite{Anderle:2021wcy}.

Theoretically, $F_\pi(Q^2)$ at low $Q^2$ can be studied using chiral perturbation theory~\cite{Gasser:1983yg} and lattice QCD~\cite{Martinelli:1987bh, Draper:1988bp, QCDSFUKQCD:2006gmg, Boyle:2008yd, JLQCD:2009ofg, Bali:2013gya, Fukaya:2014jka, Colangelo:2018mtw, Feng:2019geu, Wang:2020nbf, Gao:2021xsm, Ding:2024lfj}\footnote{The pion EMFF has recently been analyzed using the Dispersive Matrix (DM) approach~\cite{Simula:2023ujs}.}. At high $Q^2$, $F_\pi(Q^2)$ has been calculated under various approaches. Specifically, by using pQCD within the collinear factorization framework~\cite{Lepage:1979zb, Lepage:1979za, Lepage:1980fj, Efremov:1978rn, Efremov:1979qk, Duncan:1979hi, Duncan:1979ny}, it can be expressed as the following form:
\begin{eqnarray}
F_\pi(Q^2) = \int\!\!\! \int dx\,dy\,\Phi_\pi^*(x,\mu_f) T(x,y,Q^2,\mu_r^2,\mu_f^2) \Phi_\pi(y,\mu_f), \nonumber\\
\label{EQ:EMFF}
\end{eqnarray}
where $T(x, y, Q^2, \mu_r, \mu_f)$ is the perturbatively calculable hard-scattering kernel, $\Phi_\pi(y,\mu_f)$ is the nonperturbative leading-twist pion light-cone distribution amplitude (LCDA). The scale $\mu_r$ stands for the renormalization scale, and $\mu_f$ represents the factorization scale.

The perturbatively calculable hard-scattering kernel $T(x, y, Q^2, \mu_r, \mu_f)$ can be expressed as an expansion in the coupling constant $\alpha_s(\mu_r)$, i.e.,
\begin{eqnarray}
T = \frac{16\,C_F\,\pi\,\alpha_s}{Q^2}\left[T^{(0)}+T^{(1)}\frac{\alpha_s}{4\pi}+T^{(2)}\left(\frac{\alpha_s}{4\pi}\right)^2+\cdots\right].
\label{EQ:Tqcd}
\end{eqnarray}
The leading order (LO) QCD corrections have been computed more than forty years
ago~\cite{Lepage:1979za,Efremov:1979qk,Duncan:1979hi,Lepage:1980fj,Chernyak:1977fk,Farrar:1979aw,Chernyak:1980dj}. The next-to-leading order (NLO) QCD corrections have been given in Refs.\cite{Field:1981wx,Dittes:1981aw,Sarmadi:1982yg,Khalmuradov:1984ij,Braaten:1987yy,Melic:1998qr}. Recently, the next-to-next-to-leading-order (NNLO) QCD corrections have been calculated in Refs.\cite{Chen:2023byr,Ji:2024iak}.

The NNLO QCD results show that the perturbative series exhibits a slow convergence and the predictions are plagued by strong renormalization scale uncertainties for the pion EMFF~\cite{Chen:2023byr,Ji:2024iak}. It is noted that NNLO QCD predictions are determined by using the conventional scale-setting method, i.e., the renormalization scale is often set to $\mu^2_r=Q^2$ and theoretical uncertainties are evaluated by varying the same scale $\mu_r$ in an arbitrary range. This conventional procedure violates the fundamental principle of renormalization group invariance and is also inconsistent with the well-known {Gell-Mann}-Low (GM-L) method used in QED~\cite{Gell-Mann:1954yli},
it leads both to the ``renormalon" $n !$-factorial divergence~\cite{Beneke:1998ui} in the pQCD series and to the renormalization scheme-and-scale ambiguities in the pQCD predictions~\cite{Wu:2013ei,Wu:2014iba,Wu:2019mky,DiGiustino:2023jiq}.

The principle of maximum commonality (PMC)~\cite{Brodsky:2011ta, Brodsky:2011ig, Brodsky:2012rj, Brodsky:2013vpa, Mojaza:2012mf} is introduced with the aim of resolving renormalization scheme and scale ambiguities. The PMC method generalizes the BLM scale-setting procedure~\cite{Brodsky:1982gc} to all orders. PMC scales are determined by absorbing the $\beta$ terms which govern the behavior of the running $\alpha_s$ via the Renormalization Group Equations (RGE). Due to the elimination of the divergent renormalon terms, the convergence of the PMC perturbative series can be generally improved. Moreover, the PMC perturbative series becomes scale-invariant and the renormalization scale uncertainty is eliminated.

By applying the PMC method to the pion transition form factor, an improved QCD prediction without conventional renormalization scale ambiguity can be achieved~\cite{Zhou:2023ivj}. In this paper, we perform a detailed analysis for the pion EMFF $F(Q^2)$ by applying the PMCs single scale-setting approach~\cite{Shen:2017pdu}. The PMCs leads to a scale-invariant pQCD prediction and the residual scale dependence due to the unknown higher-order terms is highly suppressed.

\section{PMC scale-setting for the pion EMFF $F(Q^2)$}

The nonperturbative leading-twist pion LCDA can be expanded in the Gegenbauer polynomials with multiplicatively renormalizable coefficients,
\begin{eqnarray}
\Phi_\pi(x,\mu_f) &=& \frac{f_\pi}{2\sqrt{2N_C}}\sum_{n=0}^6a_n(\mu_f) \nonumber\\
&& \times6\,x\,(1-x)\,C_n^{3/2}\,(2x-1).
\label{EQ:Phi}
\end{eqnarray}
Here $f_\pi$ is the pion decay constant, $a_n(\mu_f)$ are the Gegenbauer moments, $C_n^{3/2}\,(2x-1)$ are the Gegenbauer polynomials, and $\sum\limits_{n=0}^6$ denotes the sum over even integers. Substituting Eq.(\ref{EQ:Phi}) into Eq.(\ref{EQ:EMFF}), the pion EMFF $F_\pi(Q^2)$ can ultimately be written as
\begin{eqnarray}
&&Q^2F_\pi(Q^2)=(e_u-e_d)\,2\,C_F\,\pi^2\,f^2_\pi \nonumber\\
&&\times\big[c_{1,0}\,a_{s}(\mu_{r})+ \big(c_{2,0} +c_{2,1}n_{f}\big)\,a_{s}^{2}(\mu_{r})\nonumber\\
&&+\big(c_{3,0}+c_{3,1}n_{f}+c_{3,2}n_{f}^{2}\big)\,a_{s}^{3}(\mu_{r}) +\mathcal{O}\big(a_{s}^{4}\big)\big],
\label{EQ:Phicij}
\end{eqnarray}
where $a_{s}(\mu_{r})=\alpha_{s}(\mu_{r})/4\pi$, $n_f$ is the number of active light-quark flavours. The pQCD coefficients up to $a_{s}^{3}$ in the $\overline{\rm MS}$ scheme were given in Refs.\cite{Chen:2023byr,Ji:2024iak}. For convenience, we put the $c_{i,j}$ coefficients in the $\overline{\rm MS}$ scheme for $\mu_{f}=Q$ in the Appendix.

After applying the PMC, its conformal series is scheme-invariant due to renormalization group invariance. However its perturbative behavior may be different for different schemes. Before applying the PMC method to the pion EMFF $F_\pi(Q^2)$, we first transform pQCD series from the $\overline{\rm MS}$ scheme to the momentum space subtraction scheme (MOM)~\cite{Celmaster:1979km,Celmaster:1979dm,Celmaster:1979xr,Celmaster:1980ji}. Unlike the $\overline{\rm MS}$ scheme, the MOM schemes are more physical schemes that carry information about the vertex at a specific momentum configuration. One may expect a higher level of convergence for expansions of the physical quantities using the physical MOM scheme. The detailed analyses for the QCD Balitsky-Fadin-Kuraev-Lipatov (BFKL) by applying the BLM/PMC method together with the use of the physical MOM scheme have already been given in Refs.\cite{Brodsky:1998kn, Zheng:2013uja, Hentschinski:2012kr, Caporale:2015uva, Ivanov:2014hpa, Celiberto:2015yba, Celiberto:2016ygs, Celiberto:2016hae, Caporale:2016zkc, Celiberto:2017ptm, Bolognino:2018oth, Celiberto:2020wpk, Celiberto:2020rxb, Celiberto:2021dzy, Celiberto:2021fdp, Celiberto:2022dyf, Hentschinski:2022xnd, Celiberto:2022rfj, Celiberto:2022gji, Egorov:2023duz}. These results show that the convergence of the perturbative QCD series can be greatly improved and more reliable theoretical predictions are obtained and they show a better agreement with the experimental data. The physical MOM scheme could be considered as the preferred renormalization scheme due to better perturbative convergence. The MOM scheme is gauge dependent, it has been shown that by taking the gauge parameter $\xi\in[-1,+1]$, such gauge dependence could be quite small~\cite{Zeng:2020lwi}. Thus we will apply the usually adopted Landau gauge with $\xi=0$ to do our calculation.

In addition, if directly calculating the pion EMFF $F_\pi(Q^2)$ in the $\overline{\rm MS}$ scheme, we shall obtain the small PMC scales that increases from $0$ to $0.2$ GeV in the range of $Q\in[0,\sqrt{20}]$ GeV. These PMC scales close to or less than the critical scale $\Lambda_{\rm QCD}$. The prediction of the pion EMFF $F_\pi(Q^2)$ can be obtained by using low-energy models for the QCD coupling constant $\alpha_s$, which however will introduce extra model dependence for the prediction. After applying the PMC, all the scheme-dependent non-conformal $\{\beta_i\}$-terms have been removed, the resultant series becomes scheme-independent. The scheme independence can also be ensured by the ``commensurate scale relations" among the pQCD approximants under various schemes~\cite{Brodsky:1994eh, Lu:1992nt}. A proper choice of physical scheme is helpful to avoid the small scale problem. For the pion EMFF $F_\pi(Q^2)$, we shall transform pQCD series from the $\overline{\rm MS}$ scheme to the physical MOM scheme.

The relation between the coupling constants in the $\overline{\rm MS}$ scheme and the MOM scheme is given by:
\begin{eqnarray}
a_s = \sum_{i=1}^3 \delta^{\rm{mom}}_{i}\,(a_s^{\rm{mom}})^i,
\label{amsMOM}
\end{eqnarray}
where the $a_s^{\rm{mom}}$ is the QCD coupling constant in the MOM scheme, and the $\delta^{\rm{mom}}_{i}$ coefficients up to $a_{s}^{3}$~\cite{Gracey:2014pba,Bednyakov:2020cdf} are given by:
\begin{eqnarray}
\delta^{\rm{mom}}_1&=&1, \\
\delta^{\rm{mom}}_2&=&-26.4925+3.4168\,n_f, \\
\delta^{\rm{mom}}_3&=&443.2412-159.9938\,n_f+15.6617\,n_f^2.
\end{eqnarray}
We transform the pion EMFF $F_\pi(Q^2)$ of Eq.(\ref{EQ:Phicij}) from the $\overline{\rm MS}$ scheme to the MOM scheme by using Eq.(\ref{amsMOM}). Thus, the pion EMFF $F_\pi(Q^2)$ in the MOM scheme can be written as
\begin{eqnarray}
&&Q^2F_\pi(Q^2)=(e_u-e_d)\,2\,C_F\,\pi^2\,f^2_\pi \nonumber\\
&&\times\big[c^{\rm{mom}}_{1,0}\,a^{\rm{mom}}_{s}(\mu_{r})+ \big(c^{\rm{mom}}_{2,0} +c^{\rm{mom}}_{2,1}n_{f}\big)\,\big(a_{s}^{\rm{mom}}(\mu_{r})\big)^2\nonumber\\
&&+\big(c^{\rm{mom}}_{3,0}+c^{\rm{mom}}_{3,1}n_{f}+c^{\rm{mom}}_{3,2}n_{f}^{2}\big)\,\big(a_{s}^{\rm{mom}}(\mu_{r})\big)^3 \nonumber\\ &&+\mathcal{O}\big((a_{s}^{\rm{mom}})^{4}\big)\big].
\label{EQ:PhicijMOM}
\end{eqnarray}
The coefficients $c^{\rm{mom}}_{i,j}$ in the MOM scheme are
\begin{eqnarray}
c^{\rm{mom}}_{1,0}&=&c_{1,0}, \\
c^{\rm{mom}}_{2,0}&=&-26.4925\,c_{1,0}+c_{2,0}, \\
c^{\rm{mom}}_{2,1}&=&3.4168\,c_{1,0}+c_{2,1}, \\
c^{\rm{mom}}_{3,0}&=&443.2412\,c_{1,0}-52.9850\,c_{2,0}+c_{3,0}, \\
c^{\rm{mom}}_{3,1}&=&-159.9938\,c_{1,0}+6.8336\,c_{2,0}-52.9850\,c_{2,1}+c_{3,1}, \nonumber\\
\\
c^{\rm{mom}}_{3,2}&=&15.6617\,c_{1,0}+6.8336\,c_{2,1}+c_{3,2}.
\end{eqnarray}

The $n_{f}$-series in Eq.(\ref{EQ:PhicijMOM}) can be unambiguously associated with the $\{\beta_{i}\}$-terms that govern the running behavior of the QCD coupling via the Renormalization Group Equation (RGE). Thus, the $n_f$ series of the pion EMFF $F_\pi(Q^2)$ is transformed into the $\{\beta_i\}$ series by using the degeneracy relation~\cite{Mojaza:2012mf},
\begin{eqnarray}
&&Q^2F_\pi(Q^2)=(e_u-e_d)\,2\,C_F\,\pi^2\,f^2_\pi \nonumber\\
&&\times\big[r^{\rm{mom}}_{1,0}\,a_{s}^{\rm{mom}}(\mu_{r}) + \big(r^{\rm{mom}}_{2,0}  + \beta_{0}r^{\rm{mom}}_{2,1}\big) \big(a_{s}^{\rm{mom}}(\mu_{r})\big)^2 \nonumber\\
&&+\big(r^{\rm{mom}}_{3,0} +\beta_{1}r^{\rm{mom}}_{2,1}+2\beta_{0}r^{\rm{mom}}_{3,1}+ \beta_{0}^{2}r^{\rm{mom}}_{3,2}\big) \big(a_{s}^{\rm{mom}}(\mu_{r})\big)^3 \nonumber\\
&&+\mathcal{O}\big((a_{s}^{\rm{mom}})^{4}\big)\big],
\label{EQ:Phirij}
\end{eqnarray}
where $\beta_{0}=11-{2\over 3}\,n_{f}$, $\beta_{1}=102-{38\over 3}\,n_{f}$. The coefficients $r^{\rm{mom}}_{i,j}$ are derived from $c^{\rm{mom}}_{i,j}(\mu_{r})$ i.e.,
\begin{eqnarray}
r^{\rm{mom}}_{1,0}&=&c^{\rm{mom}}_{1,0}, \\
r^{\rm{mom}}_{2,0}&=&c^{\rm{mom}}_{2,0}+\frac{33}{2}\,c^{\rm{mom}}_{2,1}, \\
r^{\rm{mom}}_{2,1}&=&-\frac{3}{2}\,c^{\rm{mom}}_{2,1}, \\
r^{\rm{mom}}_{3,0}&=&c^{\rm{mom}}_{3,0}+\frac{1}{4}\big(-642\,c^{\rm{mom}}_{2,1}+33\,(2\,c^{\rm{mom}}_{3,1}+33\,c^{\rm{mom}}_{3,2})\big), \nonumber\\
\\
r^{\rm{mom}}_{3,1}&=&\frac{1}{4}\big(57\,c^{\rm{mom}}_{2,1}-3\,c^{\rm{mom}}_{3,1}-99\,c^{\rm{mom}}_{3,2}\big), \\
r^{\rm{mom}}_{3,2}&=&\frac{9}{4}\,c^{\rm{mom}}_{3,2}.
\end{eqnarray}
The coefficients $r^{\rm{mom}}_{i,0}$ with $i=(1,2,3)$ are scale-invariant conformal terms, and the coefficients $r^{\rm{mom}}_{i,j}$ with $1\leq j<i\leq 3$ are non-conformal terms that should be reabsorbed into the QCD running coupling by shifting the renormalization scale.

Once all non-conformal terms are reabsorbed, a scheme-independent conformal series is obtained, that is:
\begin{eqnarray}
&&Q^2F_\pi(Q^2)=(e_u-e_d)\,2\,C_F\,\pi^2\,f^2_\pi \nonumber\\
&&\times\big[r^{\rm{mom}}_{1,0}\,a_{s}^{\rm{mom}}(Q_\star)+r^{\rm{mom}}_{2,0}\,\big(a_{s}^{\rm{mom}}(Q_\star)\big)^2 \nonumber\\
&&+r^{\rm{mom}}_{3,0}\,\big(a_{s}^{\rm{mom}}(Q_\star)\big)^3+\mathcal{O}\big((a_{s}^{\rm{mom}})^{4}\big)\big].
\label{EQ:PhirijPMC}
\end{eqnarray}
Here the $Q_{\star}$ is the PMC scale, which can be written as
\begin{eqnarray}
\ln\frac{Q^2_{\star}}{Q^2}=A_0+A_1\,a_{s}^{\rm{mom}}(Q)+\mathcal{O}\big((a_{s}^{\rm{mom}})^{2}\big),
\label{EQ:PMCscale1}
\end{eqnarray}
where
\begin{eqnarray}
A_0 &=& -{\hat{r}^{\rm{mom}}_{2,1}\over \hat{r}^{\rm{mom}}_{1,0}},  \\
A_1 &=& {2(\hat{r}^{\rm{mom}}_{2,0}\hat{r}^{\rm{mom}}_{2,1}-\hat{r}^{\rm{mom}}_{1,0}\hat{r}^{\rm{mom}}_{3,1})\over  (\hat{r}^{\rm{mom}}_{1,0})^2} \nonumber\\
&&+{((\hat{r}^{\rm{mom}}_{2,1})^2-\hat{r}^{\rm{mom}}_{1,0}\hat{r}^{\rm{mom}}_{3,2})\over (\hat{r}^{\rm{mom}}_{1,0})^2}\beta_0,
\end{eqnarray}
where $\hat{r}^{\rm{mom}}_{i,j}=r^{\rm{mom}}_{i,j}|_{\mu_r^2=Q^2}$. At present, the PMC scale $Q_{\star}$ is determined up to
next-leading logarithmic (NLL) order accuracy. The PMC scale $Q_{\star}$ is independent of any choice of scale $\mu_r$, and so are the conformal coefficients $r^{\rm{mom}}_{i,0}$, thus the resulting pQCD series is void of renormalization scale ambiguities.

\section{Numerical results and discussions}

In order to perform the numerical calculations, we adopt the following value for the pion decay constant: $f_\pi=0.131$ GeV~\cite{Chen:2023byr}. Besides, the three-loop QCD coupling constant and its $\Lambda^{\overline{\rm{MS}}}_{\rm QCD}$ scale are determined by using the world average value $\alpha_s(M_Z)=0.1180$~\cite{ParticleDataGroup:2024cfk}. The asymptotic scale in the MOM scheme is determined by $\Lambda^{\rm{mom}}_{\rm QCD}=\Lambda^{\overline{\rm{MS}}}_{\rm QCD}\,\exp[-\delta^{\rm{mom}}_2/(2\,\beta_0)]$~\cite{Celmaster:1979km}.

\begin{figure}[htb]
\centering
\includegraphics[width=0.40\textwidth]{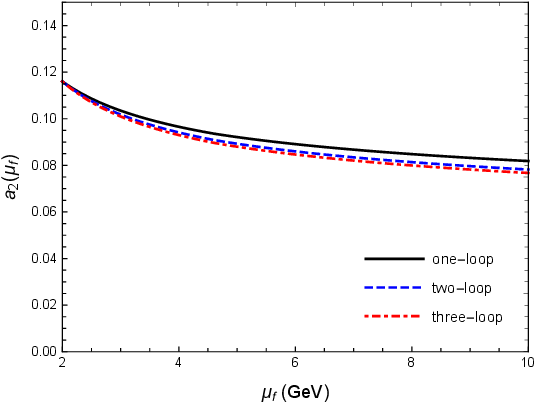}
\caption{The evolution of the Gegenbauer moment $a_2(\mu_f)$ from $a_2(2\,\rm{GeV})=0.116$ at the one-loop, two-loop and three-loop levels. }
\label{Fig:a2muf}
\end{figure}

The Gegenbauer moments $a_0(2\,\rm{GeV})=1$, $a_2(2\,\rm{GeV})=0.116^{+0.019}_{-0.020}$, $a_4(2\,\rm{GeV})=0$, and $a_6(2\,\rm{GeV})=0$ determined from lattice QCD by the RQCD Collaboration~\cite{RQCD:2019osh} is adopted as default input. We use the three-loop evolution equation~\cite{Braun:2017cih,Strohmaier:2018tjo} to evolve each Gegenbauer moment $a_n$ to any scale $\mu_f$. More explicitly, in Fig.(\ref{Fig:a2muf}), we present the evolution of the Gegenbauer moment $a_2(\mu_f)$ from $a_2(2\,\rm{GeV})=0.116$ at the one-loop, two-loop and three-loop levels. The Gegenbauer moment $a_2(\mu_f)$ shows a monotonic decrease with increasing factorization scale $\mu_f$, and the evolution shows only marginal differences between the two-loop and three-loop results. We put the three-loop evolution equation for the Gegenbauer moment $a_n(\mu_f)$ in the Appendix.

\begin{figure}[htb]
\centering
\includegraphics[width=0.40\textwidth]{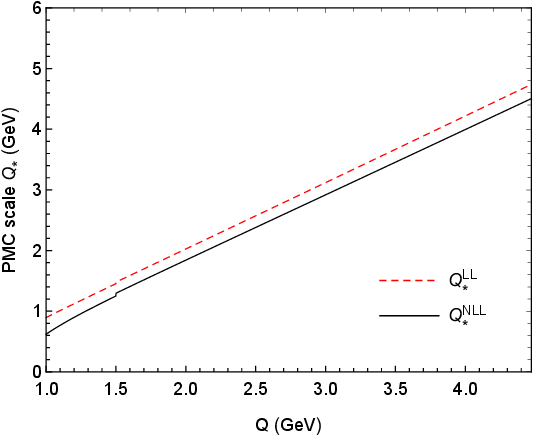}
\caption{The PMC scale $Q_{\star}$ versus $Q$ for the pion EMFF $F_\pi(Q^2)$, where the superscripts ``LL" and ``NLL" stand for the scale $Q_{\star}$ up to leading logarithmic order and next-to-leading logarithmic order accuracy, respectively.}
\label{Fig:PMCscale}
\end{figure}

The PMC scale $Q_{\star}$ versus $Q^2$ for the pion EMFF $F_\pi(Q^2)$ is shown in Fig.(\ref{Fig:PMCscale}). The PMC scale is set by the reabsorbtion of the non-conformal $\beta$ terms into the running coupling, and it is a perturbative expansion series in $\alpha_s$, as shown by Eq.(\ref{EQ:PMCscale1}). Fig.(\ref{Fig:PMCscale}) shows that the PMC scale $Q_{\star}$ monotonously increases with the increment of $Q^2$. In the low $Q^2$ region, the PMC scale $Q_{\star}$ is smaller than the momentum transfer $Q$, and the discrepancy between $Q_{\star}$ and $Q$ is small in the large $Q^2$ region. The PMC scale $Q_{\star}^{\rm{NLL}}$ determined using the NNLO QCD correction only slightly changes the PMC scale $Q_{\star}^{\rm{LL}}$ determined using the NLO QCD correction. Thus, the PMC scale shows a faster pQCD convergence, e.g., after including the NNLO QCD correction, the PMC scales slightly change down from $2.3$ GeV to $2.1$ GeV at $Q^2=5$ GeV$^2$, from $3.3$ GeV to $3.1$ GeV at $Q^2=10$ GeV$^2$, and from $4.1$ GeV to $3.9$ GeV at $Q^2=15$ GeV$^2$.

\begin{figure}[htb]
\centering
\includegraphics[width=0.40\textwidth]{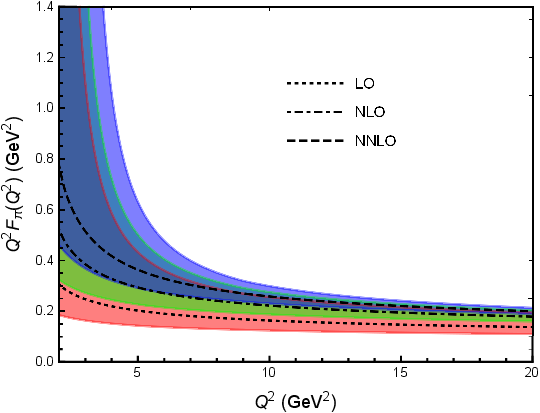}
\caption{The renormalization scale uncertainties for the pion EMFF $Q^2F_\pi(Q^2)$ using conventional scale setting, where the dotted, dotdashed and dashed lines correspond to the central values for the LO, NLO and NNLO results, respectively. The factorization scale is fixed to $\mu_f=Q$, and the renormalization scale has been varied in the range of $\mu_r\in[Q/2,2Q]$. The red, green and blue bands correspond to the LO, NLO and NNLO scale errors, respectively.}
\label{Fig:ConsRQCD}
\end{figure}

It is a common practice in literature to set the scale to $\mu_r=\mu_f=Q$, and then varying it over an arbitrary range to ascertain theoretical uncertainties. In the present analysis, we independently vary both the renormalization scale and the factorization scale to assess theoretical uncertainties. In Fig.(\ref{Fig:ConsRQCD}) we present the renormalization scale uncertainties for the pion EMFF $Q^2F_\pi(Q^2)$ using conventional scale setting, where the factorization scale is fixed to $\mu_f=Q$ and the renormalization scale is varied by $\mu_r\in[Q/2,2Q]$. We observe from Fig.(\ref{Fig:ConsRQCD}) that the NLO and NNLO corrections are positive and sizeable; even for QCD corrections up to NNLO, the conventioanl predictions are plagued by the large renormalization scale $\mu_r$ uncertainty, especially in the low $Q^2$ region.

\begin{widetext}
\begin{table} [htb]
\centering
\begin{tabular}{|c|c|c|c|c|}
\hline
 ~$Q^2$~      & ~$Q^2F_\pi(Q^2)|_{\rm{LO}}$~ & ~$Q^2F_\pi(Q^2)|_{\rm{NLO}}$~ & $Q^2F_\pi(Q^2)|_{\rm{NNLO}}$ \\
\hline
 $5 ~{\rm GeV^2}$  &~$0.2023^{+0.2504+0.0237}_{-0.0594-0.0077}$ ~&~ $0.2575^{+0.2283+0.0287}_{-0.0602-0.0071}$ ~&~ $0.3086^{+0.2920+0.0477}_{-0.0634-0.0050}$~   \\
 $10 ~{\rm GeV^2}$ &~$0.1634^{+0.0924+0.0089}_{-0.0512-0.0064}$ ~&~ $0.1956^{+0.0521+0.0079}_{-0.0520-0.0065}$ ~&~ $0.2220^{+0.0424+0.0120}_{-0.0549-0.0061}$~   \\
 $15 ~{\rm GeV^2}$ &~$0.1471^{+0.0684+0.0066}_{-0.0429-0.0051}$ ~&~ $0.1717^{+0.0331+0.0053}_{-0.0407-0.0048}$ ~&~ $0.1908^{+0.0233+0.0077}_{-0.0408-0.0043}$~   \\
\hline
\end{tabular}
\caption{Predictions for the pion EMFF $Q^2F_\pi(Q^2)$ at LO, NLO and NNLO using conventional scale setting at $Q^2=5$, $10$, $15$ GeV$^2$, where the first error is caused by varying the renormalization scale $\mu_r\in[Q/2,2Q]$, and the second error is due to the factorization scale $\mu_f\in[Q/2,2Q]$.}
\label{tab1}
\end{table}
\end{widetext}

More explicitly, we present predictions for the pion EMFF $Q^2F_\pi(Q^2)$ at LO, NLO and NNLO using conventional scale setting at $Q^2=5$, $10$, $15$ GeV$^2$ in Table \ref{tab1}. For the cause of fixing the factorization scale $\mu_f=Q$ and varying the renormalization scale $\mu_r\in[Q/2,2Q]$, the scale $\mu_r$ uncertainties at $Q^2=5$ GeV$^2$ are $[-29.4\%,+123.7\%]$, $[-23.4\%,+88.7\%]$ and $[-20.5\%,+94.6\%]$ for the LO, NLO, and NNLO predictions, respectively. At $Q^2=10$ GeV$^2$, the scale $\mu_r$ uncertainties are $[-31.3\%,+56.5\%]$, $[-26.6\%,+26.6\%]$ and $[-24.7\%,+19.1\%]$ for the LO, NLO, and NNLO predictions, respectively. At $Q^2=15$ GeV$^2$, the scale $\mu_r$ uncertainties are $[-29.1\%,+46.5\%]$, $[-23.7\%,+19.3\%]$ and $[-21.4\%,+12.2\%]$ for the LO, NLO, and NNLO predictions, respectively.

It is noted that some other scale choices have also been suggested to analyze the theoretical uncertainty for the pion EMFF $F_\pi(Q^2)$. For example, the scale choice of $\mu_r=\mu_f\in[Q/2,Q]$ has been adopted by Ref.\cite{Chen:2023byr}. In Ref.\cite{Ji:2024iak}, the renormalization scale was chosen in the range $\mu_r\in[Q/2,2Q]$, and the factorization scale was set to $\mu_f\in[Q/2,\sqrt{3}Q/2]$. For the photon-pion transition form factor, the scales choice of $\mu_r=\mu_f\in[Q/2,\sqrt{3}Q/2]$ has been suggested in Refs.\cite{Gao:2021iqq,Braun:1999uj,Agaev:2010aq}. Using this smaller scale range, a smaller scale uncertainty can be obtained.

\begin{figure}[htb]
\centering
\includegraphics[width=0.40\textwidth]{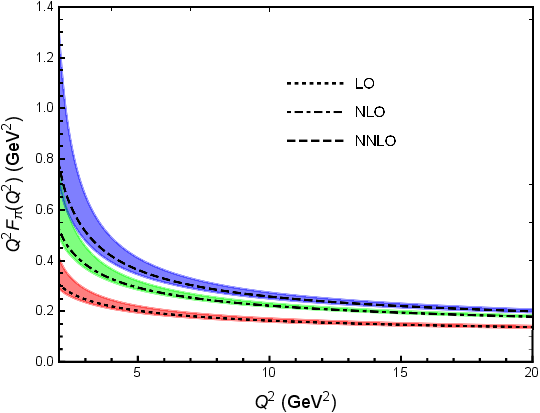}
\caption{The factorization scale uncertainties for the pion EMFF $Q^2F_\pi(Q^2)$ using conventional scale setting, where the dotted, dotdashed and dashed lines correspond to the central values for the LO, NLO and NNLO results, respectively. The renormalization scale is fixed to $\mu_r=Q$, and the factorization scale has been varied in the range: $\mu_f\in[Q/2,2Q]$. The red, green and blue bands correspond to the LO, NLO and NNLO scale errors, respectively.}
\label{Fig:ConfRQCD}
\end{figure}

By fixing the renormalization scale to $\mu_r=Q$, and varying the factorization scale $\mu_f\in[Q/2,2Q]$, the resulting theoretical predictions using conventional scale setting are presented in Fig.(\ref{Fig:ConfRQCD}). Compared to the renormalization scale $\mu_r$ uncertainty, the factorization scale $\mu_f$ uncertainty is small. At $Q^2=5$ GeV$^2$, Table \ref{tab1} shows that the scale $\mu_f$ uncertainties are $[-3.8\%,+11.7\%]$ for the LO prediction $Q^2F_\pi(Q^2)|_{\rm{LO}}$, $[-2.8\%,+11.1\%]$ for the NLO prediction $Q^2F_\pi(Q^2)|_{\rm{NLO}}$, and $[-1.6\%,+15.5\%]$ for the NNLO prediction $Q^2F_\pi(Q^2)|_{\rm{NNLO}}$. At $Q^2=10$ GeV$^2$, the scale $\mu_f$ uncertainties are $[-3.9\%,+5.4\%]$ for the $Q^2F_\pi(Q^2)|_{\rm{LO}}$, $[-3.3\%,+4.0\%]$ for the $Q^2F_\pi(Q^2)|_{\rm{NLO}}$, and $[-2.7\%,+5.4\%]$ for the $Q^2F_\pi(Q^2)|_{\rm{NNLO}}$. At $Q^2=15$ GeV$^2$, the scale $\mu_f$ uncertainties are $[-3.5\%,+4.5\%]$, $[-2.8\%,+3.1\%]$ and $[-2.3\%,+4.0\%]$ for the $Q^2F_\pi(Q^2)|_{\rm{LO}}$, $Q^2F_\pi(Q^2)|_{\rm{NLO}}$ and $Q^2F_\pi(Q^2)|_{\rm{NNLO}}$, respectively.

At LO, the factorization scale $\mu_f$ uncertainty stems solely from the Gegenbauer moment $a_n(\mu_f)$. Beyond LO, however, this uncertainty arises not only from the Gegenbauer moment $a_n(\mu_f)$ but also from the logarithmic term $\ln\frac{\mu_f^2}{Q^2}$ in the perturbative coefficients. Although the Gegenbauer moment $a_n(\mu_f)$ decrease monotonically with increasing factorization scale $\mu_f$, this dependence is not particularly strong, as demonstrated in Fig.(\ref{Fig:a2muf}). It is observed that the sizable coefficients associated with the logarithmic term $\ln\frac{\mu_f^2}{Q^2}$ beyond LO are responsible for the relatively large factorization scale uncertainty. Consequently, the uncertainties at NLO and NNLO are larger than those at LO.

\begin{figure}[htb]
\centering
\includegraphics[width=0.40\textwidth]{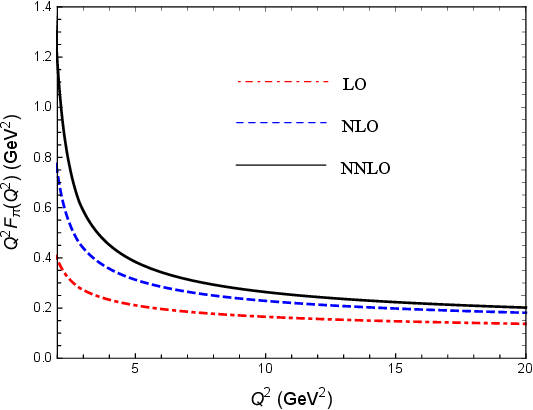}
\caption{The renormalization scale uncertainties for the pion EMFF $Q^2F_\pi(Q^2)$ using PMC scale setting, where the dotted, dotdashed and dashed lines correspond to the LO, NLO and NNLO results, respectively. The factorization scale is fixed to $\mu_f=Q$. The PMC prediction of pion EMFF $Q^2F_\pi(Q^2)$ is independent of the scale $\mu_r$. }
\label{Fig:PMCsRQCD}
\end{figure}

In the case of PMC scale setting, the renormalization scale uncertainties are eliminated. We present the renormalization scale uncertainties for the pion EMFF $Q^2F_\pi(Q^2)$ using PMC scale setting in Fig.(\ref{Fig:PMCsRQCD}). At $Q^2=5$ GeV$^2$, the scale $\mu_r$-invariant results are $Q^2F_\pi(Q^2)|_{\rm{LO}}=0.2108$, $Q^2F_\pi(Q^2)|_{\rm{NLO}}=0.3126$, and $Q^2F_\pi(Q^2)|_{\rm{NNLO}}=0.3839$ for LO, NLO and NNLO predictions, respectively. At $Q^2=10$ GeV$^2$, the scale $\mu_r$-invariant results are $Q^2F_\pi(Q^2)=0.1652$, $0.2285$, and $0.2637$ for LO, NLO and NNLO predictions, respectively. At $Q^2=15$ GeV$^2$, the scale $\mu_r$-invariant results are $Q^2F_\pi(Q^2)=0.1474$, $0.1981$, and $0.2235$ for LO, NLO and NNLO predictions, respectively. Compared to the conventional results obtained for $\mu_r=\mu_f=Q$, the PMC results are increased to a certain degree, especially, in the low $Q^2$ region, where the PMC scale is small, leading to a large QCD correction.

\begin{figure}[htb]
\centering
\includegraphics[width=0.40\textwidth]{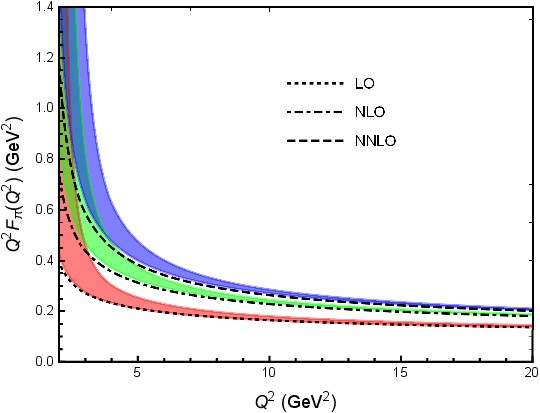}
\caption{The factorization scale uncertainties for the pion EMFF $Q^2F_\pi(Q^2)$ using PMC scale setting, where the dotted, dotdashed and dashed lines correspond to the central values for the LO, NLO and NNLO results, respectively. The factorization scale has been varied in the range: $\mu_f\in[Q/2,2Q]$. The red, green and blue bands correspond to the LO, NLO and NNLO scale errors, respectively. }
\label{Fig:PMCfRQCD}
\end{figure}

\begin{table} [htb]
\centering
\begin{tabular}{|c|c|c|c|c|}
\hline
 ~$Q^2$~      & ~$Q^2F_\pi(Q^2)|_{\rm{LO}}$~ & ~$Q^2F_\pi(Q^2)|_{\rm{NLO}}$~ & $Q^2F_\pi(Q^2)|_{\rm{NNLO}}$ \\
\hline
 $5 ~{\rm GeV^2}$  &~$0.2108^{+0.0572}_{-0.0003}$ ~&~ $0.3126^{+0.0869}_{+0.0106}$ ~&~ $0.3839^{+0.1265}_{+0.0163}$~   \\
 $10 ~{\rm GeV^2}$ &~$0.1652^{+0.0153}_{-0.0036}$ ~&~ $0.2285^{+0.0169}_{-0.0002}$ ~&~ $0.2637^{+0.0218}_{-0.0005}$~   \\
 $15 ~{\rm GeV^2}$ &~$0.1474^{+0.0101}_{-0.0029}$ ~&~ $0.1981^{+0.0098}_{-0.0002}$ ~&~ $0.2235^{+0.0123}_{-0.0006}$~   \\
\hline
\end{tabular}
\caption{Predictions for the pion EMFF $Q^2F_\pi(Q^2)$ at LO, NLO and NNLO using PMC scale setting for $Q^2=5$, $10$, $15$ GeV$^2$, where the error is due to the factorization scale $\mu_f\in[Q/2,2Q]$. The PMC predictions for the pion EMFF $Q^2F_\pi(Q^2)$ is independent of the renormalization scale $\mu_r$. }
\label{tab2}
\end{table}

The factorization scale uncertainty exists even for conformal theories. This kind of uncertainty can be resolved by matching the perturbative predictions with the nonperturbative boundstate dynamics~\cite{Brodsky:2014yha}. After applying PMC scale setting, the renormalization scale uncertainty is eliminated, while factorization scale uncertainties remain and they are shown in Fig.(\ref{Fig:PMCfRQCD}). More explicitly, we present predictions for the pion EMFF $Q^2F_\pi(Q^2)$ at LO, NLO and NNLO using PMC scale setting for $Q^2=5$, $10$, $15$ GeV$^2$ in Table \ref{tab2}. It shows that at $Q^2=5$ GeV$^2$, the scale $\mu_f$ uncertainties are $[-0.1\%,+27.1\%]$, $[+3.4\%,+27.8\%]$, and $[+4.2\%,+32.9\%]$ for the $Q^2F_\pi(Q^2)|_{\rm{LO}}$, $Q^2F_\pi(Q^2)|_{\rm{NLO}}$, and $Q^2F_\pi(Q^2)|_{\rm{NNLO}}$, respectively. At $Q^2=10$ GeV$^2$, the scale $\mu_f$ uncertainties are $[-2.2\%,+9.3\%]$, $[-0.08\%,+7.4\%]$, and $[-0.2\%,+8.3\%]$ for the $Q^2F_\pi(Q^2)|_{\rm{LO}}$, $Q^2F_\pi(Q^2)|_{\rm{NLO}}$, and $Q^2F_\pi(Q^2)|_{\rm{NNLO}}$, respectively. At $Q^2=15$ GeV$^2$, the scale $\mu_f$ uncertainties are $[-2.0\%,+6.9\%]$, $[-0.1\%,+5.0\%]$, and $[-0.3\%,+5.5\%]$ for the $Q^2F_\pi(Q^2)|_{\rm{LO}}$, $Q^2F_\pi(Q^2)|_{\rm{NLO}}$, and $Q^2F_\pi(Q^2)|_{\rm{NNLO}}$, respectively.

In the case of conventional scale setting, the quality of the convergence of the pQCD series depends on the choice of renormalization scale, and one cannot decide whether the poor convergence is an intrinsic property of the pQCD series or is simply due to the improper choice of the renormalization scale. After applying the PMC, the convergence of the pQCD series does not depend on the choice of renormalization scale and is not greatly improved for the pion EMFF $Q^2F_\pi(Q^2)$. This is caused by the presence of a large conformal term, indicating a higher QCD calculation is important. Simply varying the scale in $\alpha_s$ to estimate the unknown higher-order QCD contributions is not applicable to the PMC predictions, since the PMC scales are determined unambiguously by the non-conformal $\beta$ terms, and varying the PMC scales would break the renormalization group invariance and then lead to an unreliable prediction~\cite{Wu:2014iba}.

The Pad$\acute{e}$ approximation approach (PAA)~\cite{Basdevant:1972fe, Samuel:1992qg, Samuel:1995jc} offers a feasible conjecture that yields the unknown $(n+1)_{\rm th}$-order contributions from the given $n_{\rm th}$-order perturbative series. We adopt the PAA approach to estimate the unknown higher-order QCD contributions and a $[N/M]$-type approximant $\rho_n^{[N/M]}$ for observable $\rho_n=\sum_{i=0}^{n(\geq 1)}C_i x^i$ is defined as
\begin{eqnarray}
\rho^{[N/M]}_n &=& \frac{b_0+b_1 x + \cdots + b_N x^N} {1 + c_1 x + \cdots + c_M x^M} \nonumber \\
               &=& \sum_{i=0}^{n} C_i x^i + C_{n+1} \; x^{n+1} +\cdots,
\end{eqnarray}
where $M\geq 1$ and $N+M=n$. The known coefficients $C_{i(\leq n)}$ determine the parameters $b_{i\in[0,N]}$ and $c_{j\in[1,M]}$, which inversely predicts a reasonable value for the next unknown coefficient $C_{n+1}$~\cite{Du:2018dma}. By applying the PAA approach to the scale $\mu_r$-invariant PMC conformal series, the unknown NNNLO coefficient is $r_{4,0}^{\rm PAA}=(-r_{2,0}^3+2r_{1,0}r_{2,0}r_{3,0})/r_{1,0}^2$ for $[N/M]=[0/2]$, and thus the uncertainty from unknown higher-order contributions could be estimated by $\pm r_{4,0}^{\rm PAA}\,a^{4}_s(Q_\star)$.

\begin{figure}[htb]
\centering
\includegraphics[width=0.40\textwidth]{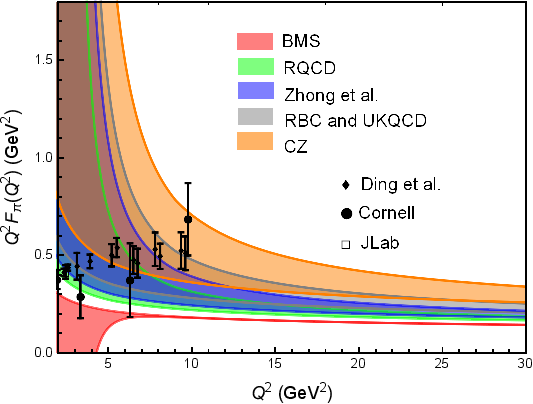}
\caption{Theoretical predictions for the pion EMFF $Q^2F_\pi(Q^2)$ using conventional scale setting with the five different pion LCDA mentioned in the text, where the shaded bands represent the conventional scale uncertainty for $\mu_r=\mu_f\in[Q/2,2Q]$. The experimental measurements from JLab~\cite{JeffersonLab:2008jve} and Cornell~\cite{Bebek:1977pe}, and the lattice QCD predictions~\cite{Ding:2024lfj} are presented for comparison. }
\label{Fig:Q2FConv}
\end{figure}

\begin{figure}[htb]
\centering
\includegraphics[width=0.40\textwidth]{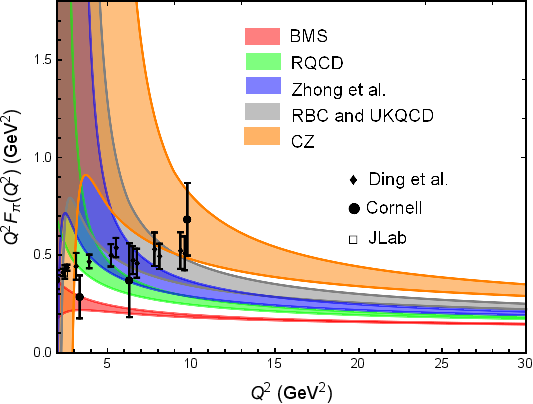}
\caption{Theoretical predictions for the pion EMFF $Q^2F_\pi(Q^2)$ using PMC scale setting with the five different pion LCDA mentioned in the text, where the shaded bands represent the uncertainty that combines the contributions from the factorization scale $\mu_f\in[Q/2,2Q]$ and the estimation of unknown higher-order contributions by $\pm r_{4,0}^{\rm PAA}\,a^{4}_s(Q_\star)$. The experimental measurements from JLab~\cite{JeffersonLab:2008jve} and Cornell~\cite{Bebek:1977pe}, and the lattice QCD predictions~\cite{Ding:2024lfj} are presented for comparison. }
\label{Fig:Q2FPMC}
\end{figure}

In addition to the uncertainty of the unknown higher-order QCD contributions, the renormalization and factorization scales uncertainties, the pion LCDA is a key nonperturative input parameter, which has been extensively investigated from different approaches. Besides the lattice QCD result from the RQCD Collaboration~\cite{RQCD:2019osh}, the Gegenbauer moments $a_2(1\,\rm{GeV})=0.56$ determined from CZ parametrization in the early days~\cite{Chernyak:1983ej}, $a_2(2\,\rm{GeV})=0.233\pm0.065$ determined from lattice QCD by the RBC and UKQCD Collaborations~\cite{Arthur:2010xf}, $a_2(2\,\rm{GeV})=0.149^{+0.052}_{-0.043}$, $a_4(2\,\rm{GeV})=-0.096^{+0.063}_{-0.058}$ extracted from QCD sum rules with non-local condensates (BMS approach)~\cite{Bakulev:2001pa,Mikhailov:2016klg,Stefanis:2020rnd}, and $a_2(1\,\rm{GeV})=0.206\pm0.038$, $a_4(1\,\rm{GeV})=0.047\pm0.011$ obtained by using the QCD sum rules in the framework of background field theory~\cite{Zhong:2021epq} are also adopted as input parameters. For all the Gegenbauer moments mentioned above, $a_0=1$, and the remaining unspecified ones are $0$. More explicitly, we present the theoretical predictions for the pion EMFF $Q^2F_\pi(Q^2)$ using conventional and PMC scale settings with these five different pion LCDA in Figs. (\ref{Fig:Q2FConv}) and (\ref{Fig:Q2FPMC}). The conventional predictions are plagued by large scale uncertainties, especially at large momentum transfer $Q^2$ region compared to the PMC predictions. These sizeable uncertainties of the conventional predictions prevent a precise study of the determination of the pion light-cone distribution amplitude and a detailed comparison with experimental results.

While, by introducing the PMC the renormalization scale uncertainties can be eliminated and the precision of theoretical predictions can be improved, especially at large momentum transfer $Q^2$ region. In the low-$Q^2$ region, the large QCD corrections cause the breakdown of perturbative QCD, rendering the PAA method for estimating unknown higher orders ineffective, leading to significant errors. Figure (\ref{Fig:Q2FPMC}) shows that PMC predictions agree with the experimental data in the low-$Q^2$ region when the Gegenbauer moments are determined from BMS approach~\cite{Bakulev:2001pa,Mikhailov:2016klg,Stefanis:2020rnd}; when the Gegenbauer moments are taken from CZ~\cite{Chernyak:1983ej}, RQCD~\cite{RQCD:2019osh}, RBC and UKQCD~\cite{Arthur:2010xf}, and Zhong et al.~\cite{Zhong:2021epq}, we find satisfactory agreement between PMC predictions and the experimental data in the $Q^2$ region from $4$ to $10$ GeV$^2$.

\section{Summary}

The dynamics of the $\pi$ mesons plays a crucial role in the development of QCD, and careful study of its EMFF \(F_\pi(Q^2)\) is of great importance for understanding the strong interactions. When applying the conventional scale setting, the significant uncertainty in the pion EMFF \(F_\pi(Q^2)\) is dominated by the choice of the renormalization scale. This sizeable uncertainty affects the conventional predictions preventing precise study of the determination of the pion LCDA and detailed comparison with experimental results. Unlike the conventional scale-setting method, the PMC method eliminates uncertainties associated with renormalization scheme and scale ambiguities. In this paper, we provide a detailed PMC analysis of the pion EMFF \(Q^2F_\pi(Q^2)\) in the physical MOM scheme. A more precise pQCD prediction, which is free of conventional renormalization scale ambiguity, for the pion EMFF \(F_\pi(Q^2)\) has been achieved. Such improved pQCD prediction will be highly beneficial for precise study of the determination of the pion LCDA as well as for detailed comparison with experimental data.

\hspace{1cm}

{\bf Acknowledgements}: The authors would like to thank Tao Zhong and Long-Bin Chen for helpful discussions. This work was supported in part by the Natural Science Foundation of China under Grant No.12265011, No.12175025, No.12347101 and No.12275036; by the Project of Guizhou Provincial Department of Science and Technology under Grant No.YQK[2023]016, No.ZK[2023]141 and No.CXTD[2025]030; by the Hunan Provincial Natural Science Foundation with Grant No.2024JJ3004, YueLuShan Center for Industrial Innovation (2024YCII0118); by the Natural Science Foundation of Sichuan Province under Grant No.2024NSFSC1367; and the Natural Science Foundation of Chongqing under Grants No.cstc2021jcyj-msxmX0681.

\appendix

\section{The coefficients $c_{i,j}$}

Based on the calculation of Refs.\cite{Chen:2023byr,Ji:2024iak}, setting the factorization scale as $\mu_f=Q$ and expanding the Gegenbauer up to the sixth conformal moment, the coefficients $c_{i,j}$ in Eq.(\ref{EQ:Phicij}) in the $\overline{\rm MS}$ scheme up to $a_{s}^{3}$ are
\begin{widetext}
\begin{eqnarray}
c_{1,0}&=&5.4199a_0^2 + 5.4199a_2^2 + 5.4199a_4^2 + 10.8398 a_4a_6 + 5.4199a_6^2 + a_2(10.8398a_4 + 10.8398a_6) \nonumber\\
&& + a_0(10.8398a_2 + 10.8398a_4 + 10.8398a_6);  \\ \nonumber\\
c_{2,0}&=&193.3102a_0^2 + 560.3497a_2^2 + 1369.2194a_2a_4 + 822.0682a_4^2 + 1550.5995a_2a_6 + 1843.0916a_4a_6 + 1026.8759a_6^2 \nonumber\\
&& + a_0(688.1996a_2 + 878.2097a_4 + 1020.7008a_6) + \left(59.6190a_0^2 + 59.6190a_2^2 + 119.2381a_2a_4 + 59.6190a_4^2 \right. \nonumber\\
&& \left. + 119.2381a_2a_6 + 119.2381a_4a_6 + 59.6190a_6^2 + a_0(119.2381a_2 + 119.2381a_4 + 119.2381a_6)\right) \ln\frac{\mu_r^2}{Q^2}, \\
c_{2,1}&=&-16.8620a_0^2 - 31.9173a_2^2 - 70.6998a_2a_4 - 38.7825a_4^2 + a_0(-48.7792a_2 - 55.6444a_4 - 60.2299a_6) \nonumber\\
&& - 75.2852a_2a_6 - 82.1504a_4a_6 - 43.3679a_6^2 + \left(-3.6133a_0^2 - 3.6133a_2^2 - 7.2266a_2a_4 - 3.6133a_4^2 \right. \nonumber\\
&& \left. + a_0(-7.2266a_2 - 7.2266a_4 - 7.2266a_6) - 7.2266a_2a_6 - 7.2266a_4a_6 - 3.6133a_6^2\right)\ln\frac{\mu_r^2}{Q^2}; \\ \nonumber\\
c_{3,0}&=&9900.1794a_0^2 + 54482.8691a_2^2 + 152766.0088a_2a_4 + 104539.9677a_4^2 + 189111.4788a_2a_6 + 255052.1389a_4a_6 \nonumber\\
&& + 154179.9730a_6^2 + a_0(49210.9309a_2 + 72957.3863a_4 + 93512.9011a_6) + \left(4805.6562a_0^2 + 12880.5243a_2^2 \right.\nonumber\\
&& \left. + 31228.4895a_2a_4 + 18638.3313a_4^2 + 35218.8519a_2a_6 + 41653.6769a_4a_6 + 23144.1015a_6^2 \right.\nonumber\\
&& \left. + a_0(16246.0538a_2 + 20426.2761a_4 + 23561.0789a_6)\right)\ln\frac{\mu_r^2}{Q^2} + \left(655.8095a_0^2 + 655.8095a_2^2 \right.\nonumber\\
&& \left. + 1311.6189a_2a_4 + 655.8095a_4^2 + 1311.6189a_2a_6 + 1311.6189a_4a_6 + 655.8095a_6^2 \right. \nonumber\\
&& \left. + a_0(1311.6189a_2 + 1311.6189a_4 + 1311.6189a_6)\right)\ln^2\frac{\mu_r^2}{Q^2}, \\
c_{3,1}&=&-1822.71166a_0^2 - 6873.8606a_2^2 - 17803.9941a_2a_4 - 11305.2700a_4^2 + a_0(-7455.3656a_2 - 10122.2088a_4 \nonumber\\
&& - 12259.0011a_6) - 20951.6879a_2a_6 - 26296.9153a_4a_6 - 15184.0479a_6^2 + \left(-697.3621a_0^2 - 1517.9650a_2^2 \right. \nonumber\\
&& \left. - 3518.3249a_2a_4 - 2017.9579a_4^2 + a_0(-2128.0467a_2 - 2532.4284a_4 - 2823.2957a_6) - 3861.0443a_2a_6 \right. \nonumber\\
&& \left. - 4402.0686a_4a_6 - 2391.9141a_6^2\right)\ln\frac{\mu_r^2}{Q^2} + \left(-79.4921a_0^2 - 79.4921a_2^2 - 158.9841a_2a_4 - 79.4921a_4^2 \right. \nonumber\\
&& \left. + a_0(-158.9841a_2 - 158.9841a_4 - 158.9841a_6) - 158.9841a_2a_6 - 158.9841a_4a_6 - 79.4921a_6^2\right)\ln^2\frac{\mu_r^2}{Q^2}, \\
c_{3,2}&=& 58.4815a_0^2 + 188.7936a_2^2 + 462.2824a_2a_4 + 277.8368a_4^2 + 523.8835a_2a_6 + 623.0828a_4a_6 + 347.1857a_6^2 \nonumber\\
&& + a_0(226.3650a_2 + 291.9901a_4 + 340.8538a_6) + (22.4826a_0^2 + 42.5564a_2^2 + 94.2663a_2a_4 + 51.7100a_4^2 \nonumber\\
&& + 100.3802a_2a_6 + 109.5339a_4a_6 + 57.8239a_6^2 + a_0(65.0390a_2 + 74.1926a_4 + 80.3065a_6))\ln\frac{\mu_r^2}{Q^2} \nonumber\\
&& + (2.4089a_0^2 + 2.4089a_2^2 + 4.8177a_2a_4 + 2.4089a_4^2 + 4.8177a_2a_6 + 4.8177a_4a_6 + 2.4089a_6^2 \nonumber\\
&& + a_0(4.8177a_2 + 4.8177a_4 + 4.8177a_6))\ln^2\frac{\mu_r^2}{Q^2}.
\end{eqnarray}
\end{widetext}

\section{The three-loop evolution equation for the Gegenbauer moment $a_n(\mu_f)$}

The three-loop evolution equation to evolve Gegenbauer moment $a_n$ evaluated at $\mu_0$ to any scale $\mu_f$ can be written as~\cite{Braun:2017cih,Strohmaier:2018tjo}
\begin{widetext}
\begin{eqnarray}
a_n(\mu_f)&=&a_n(\mu_0)\biggl[\left(\frac{a_{s}(\mu_{0})}{a_{s}(\mu_{f})}\right)^{-\frac{\gamma^{(1)}_{n}}{2\beta_{0}}}+a_{s}(\mu_{f})\biggl(\left(\frac{a_{s}(\mu_{0})}{a_{s}(\mu_{f})}\right)^{-\frac{\gamma^{(1)}_{n}}{2\beta_{0}}}\mathcal{A}^{(1)}_n+\sum_{k=0}^{n}\left(\frac{a_{s}(\mu_{0})}{a_{s}(\mu_{f})}\right)^{-\frac{\gamma^{(1)}_{k}}{2\beta_{0}}}\mathcal{B}^{(1)}_{nk}\biggl) \nonumber\\
&&+a^2_{s}(\mu_{f})\biggl(\left(\frac{a_{s}(\mu_{0})}{a_{s}(\mu_{f})}\right)^{-\frac{\gamma^{(1)}_{n}}{2\beta_{0}}}\mathcal{A}^{(2)}_n+\sum_{k=0}^{n}\left(\frac{a_{s}(\mu_{0})}{a_{s}(\mu_{f})}\right)^{-\frac{\gamma^{(1)}_{k}}{2\beta_{0}}}\left(\mathcal{B}^{(2)}_{nk}+\mathcal{B}^{(1)}_{nk}\mathcal{A}^{(1)}_{k}\right)\biggl)\biggl].
\end{eqnarray}
\end{widetext}
The matrix elements $\mathcal{A}^{(1)}_n$, $\mathcal{A}^{(2)}_n$, $\mathcal{B}^{(1)}_{nk}$ and $\mathcal{B}^{(2)}_{nk}$ are explicitly given in Eqs.(7.24) and (7.28) of Ref.~\cite{Strohmaier:2018tjo}, and are expressed in terms of the $\beta$-functions and the anomalous dimensions. As an example, we consider the commonly used Gegenbauer moment $a_2(\mu_f)$ and present the following three-loop evolution equation,
\begin{widetext}
\begin{eqnarray}
a_2(\mu_f)&=&a_2(\mu_0)\left(\frac{a_{s}(\mu_{0})}{a_{s}(\mu_{f})}\right)^{-\frac{\gamma^{(1)}_2}{2\beta_{0}}}\biggl[1 + \frac{1}{72}\biggl(\frac{36(a_s(\mu_0)-a_s(\mu_f))(\beta_1\gamma^{(1)}_2-\beta_0\gamma^{(2)}_2)}{\beta_0^2} + \frac{9}{\beta_0^4}(a_s(\mu_0)-a_s(\mu_f)) \nonumber\\
&&\times\biggl((a_s(\mu_0)-a_s(\mu_f))(\beta_1\gamma^{(1)}_2-\beta_0\gamma^{(2)}_2)^2 + 2(a_s(\mu_0)+a_s(\mu_f))\beta_0^2(\beta_1^2\gamma^{(1)}_2+\beta_1\gamma^{(2)}_2-\beta_0(\beta_2\gamma^{(1)}_2+\gamma^{(3)}_2))\biggl) \nonumber\\
&&-\frac{28}{a_2(\mu_0)(2\beta_0-\gamma^{(1)}_2)}\biggl(a_s(\mu_0)-\left(\frac{a_{s}(\mu_{0})}{a_{s}(\mu_{f})}\right)^{\frac{\gamma^{(1)}_2}{2\beta_{0}}}a_s(\mu_f)\biggl)\gamma^{(2)}_{20} + \frac{14}{a_2(\mu_0)\beta_0^2(8\beta_0^2-6\beta_0\gamma^{(1)}_2+(\gamma^{(1)}_2)^2)}\nonumber\\
&&\times2\left(\frac{a_{s}(\mu_{0})}{a_{s}(\mu_{f})}\right)^{\frac{\gamma^{(1)}_2}{2\beta_{0}}}a_s(\mu_f)^2\beta_0+a_s(\mu_0)(2a_s(\mu_0)\beta_0-4a_s(\mu_f)\beta_0-a_s(\mu_0)\gamma^{(1)}_2+a_s(\mu_f)\gamma^{(1)}_2) \nonumber\\
&&(-\beta_1\gamma^{(1)}_2+\beta_0\gamma^{(2)}_2)\gamma^{(2)}_{20} +\frac{28}{a_2(\mu_0)\beta_0(4\beta_0-\gamma^{(1)}_2)}\biggl(a_s(\mu_0)^2-\left(\frac{a_{s}(\mu_{0})}{a_{s}(\mu_{f})}\right)^{\frac{\gamma^{(1)}_2}{2\beta_{0}}}a_s(\mu_f)^2\biggl)(\beta_1\gamma^{(2)}_{20}-\beta_0\gamma^{(3)}_{20})\biggl)\biggl].\nonumber\\
\end{eqnarray}
\end{widetext}
where the $\beta_{0}$, $\beta_{1}$ and $\beta_{2}$ are the one-loop, two-loop and three-loop $\beta$-function, respectively,
\begin{eqnarray}
\beta_{0}&=&11-\frac{2}{3}n_f, \nonumber\\
\beta_{1}&=&102-\frac{38}{3}n_f, \nonumber\\
\beta_{2}&=&\frac{2857}{2} - \frac{5033}{18}n_f + \frac{325}{54}n_f^2.
\end{eqnarray}
The matrix elements $\gamma^{(1)}_{n}$, $\gamma^{(2)}_{n}$ and $\gamma^{(3)}_{n}$ are the one-loop, two-loop and three-loop anomalous dimensions and can be written as~\cite{Braun:2017cih,Strohmaier:2018tjo},
\begin{eqnarray}
\gamma^{(1)}_2&=&\frac{100}{9}, \nonumber\\
\gamma^{(2)}_2&=&\frac{34450}{243}-\frac{830}{81}n_f, \nonumber\\
\gamma^{(3)}_2&=&\frac{64486199}{26244}+\frac{2200}{81}\zeta_3-\left(\frac{967495}{4374}+\frac{4000}{27}\zeta_3\right)n_f \nonumber\\
&& -\frac{2569}{729}n_f^2, \nonumber\\
\gamma^{(2)}_{20}&=&\frac{260}{9}-\frac{40}{9}n_f, \nonumber\\
\gamma^{(3)}_{20}&=&\frac{49024}{81}-\frac{28700}{243}n_f+\frac{236}{81}n_f^2. \nonumber\\
\end{eqnarray}

\end{document}